\documentclass[a4paper]{article}

\usepackage[english]{babel}
\usepackage[utf8]{inputenc}
\usepackage{amsmath}
\usepackage{graphicx}
\usepackage[colorinlistoftodos]{todonotes}

\usepackage{url,hyperref,lineno,microtype}
\usepackage{amsfonts,amssymb}
\usepackage{dsfont}
\usepackage{authblk}
\usepackage{booktabs}
\usepackage[font=small,labelfont=bf]{caption}

\newcommand{\f}{\boldsymbol{\phi}}
\newcommand{\g}{\mathbf{g}}

\title{Modeling the dynamics of human brain activity with recurrent neural networks}

\author{Umut Güçlü}
\author{Marcel A. J. van Gerven}
\affil{Radboud University, Donders Institute for Brain, Cognition and Behaviour, Nijmegen, the Netherlands}

\date{}

\begin{document}
\maketitle

\begin{abstract}
Encoding models are used for predicting brain activity in response to sensory stimuli with the objective of elucidating how sensory information is represented in the brain. Encoding models typically comprise a nonlinear transformation of stimuli to features (feature model) and a linear transformation of features to responses (response model). While there has been extensive work on developing better feature models, the work on developing better response models has been rather limited. Here, we investigate the extent to which recurrent neural network models can use their internal memories for nonlinear processing of arbitrary feature sequences to predict feature-evoked response sequences as measured by functional magnetic resonance imaging. We show that the proposed recurrent neural network models can significantly outperform established response models by accurately estimating long-term dependencies that drive hemodynamic responses. The results open a new window into modeling the dynamics of brain activity in response to sensory stimuli.
\end{abstract}

\section{Introduction}

Encoding models~\cite{Naselaris2011} are used for predicting brain activity in response to naturalistic stimuli~\cite{Felsen2005} with the objective of understanding how sensory information is represented in the brain. Encoding models typically comprise two main components. The first component is a feature model that nonlinearly transforms stimuli to features. The second component is a response model that linearly transforms features to responses. While encoding models have been successfully used to characterize the relationship between stimuli in different modalities and responses in different brain regions, their performance usually fall short of the expected performance of the true encoding model given the noise in the analyzed data (noise ceiling). This means that there usually is unexplained variance in the analyzed data that can be explained solely by improving the encoding models.

One way to reach the noise ceiling is the development of better feature models. Recently, there has been extensive work in this direction. One example is the use of convolutional neural network representations of natural images or natural movies to explain low-, mid- and high-level representations in different brain regions along the ventral~\cite{Yamins2014,Cadieu2014,Khaligh-Razavi2014,Agrawal2014,Guclu2015a,Cichy2016} and dorsal streams~\cite{Guclu2015c} of the human visual system. Another example is the use of manually constructed or statistically estimated representations of words and phrases to explain the semantic representations in different brain regions~\cite{Mitchell2008,Huth2012,Murphy2012,Fyshe2013,Guclu2015b,Nishida2015}.

Another way to reach the noise ceiling is the development of better response models. Unlike the progress in developing better feature models, the progress in developing better response models has been relatively slow. Standard response models linearly transform features to responses. In reality, this transformation is much more complex because of various temporal dependencies that are caused by neurovascular coupling~\cite{Logothetis2004,Norris2006} and other more elusive cognitive factors. While these models can account for the temporal dependencies that are caused by neurovascular coupling to some extent by using analytically derived~\cite{Friston1998} or statistically estimated~\cite{Dale1999,Glover1999} hemodynamic response functions (HRFs), they largely ignore the remaining temporal dependencies since either they are computationally inefficient to take into account or their underlying factors are difficult to identify. Recent promising work in this direction has been on the derivation and estimation of more accurate HRFs. For example,~\cite{Aquino2014} has shown that accurate HRFs can be analytically derived from physiology, and~\cite{Pedregosa2015} has shown that accurate HRFs can be statistically estimated from data. Note that while the methods for statistically estimating HRFs are particularly suited to use in block designs and event related designs, they are not straightforward to use in continuous designs.

Here, our objective is to develop a response model that can be trained end to end, captures temporal dependencies and processes arbitrary input sequences. To this end, we use recurrent neural networks (RNNs) as response models in the encoding framework. Recently, RNNs in general and two RNN variants - long short-term memory~\cite{Hochreiter1997} and gated recurrent units~\cite{Cho2014} - in particular have been shown to be extremely successful in various tasks that involve processing of arbitrary input sequences such as handwriting recognition~\cite{Graves2009,Graves2013}, language modeling~\cite{Sutskever2011,Graves2013}, machine translation~\cite{Cho2014} and speech recognition~\cite{Sak2014}. These models use their internal memories to capture the temporal dependencies that are informative about solving the task at hand. If these models can be used as response models in the encoding framework, it will open a new window into modeling brain activity in response to sensory stimuli. While the use of RNNs in this setting has been proposed a number of times \cite{Guclu2015a,Guclu2015c,Kriegeskorte2015,Yamins2016a,Yamins2016b}, an in depth analysis of this approach is yet to be performed.

We test this approach by comparing how well a family of recurrent neural network models and a family of ridge regression models can predict blood-oxygen-level dependent (BOLD) hemodynamic responses to high-level and low-level features of natural movies. We show that the proposed recurrent neural network models can significantly outperform the standard ridge regression models and accurately estimate hemodynamic response functions by capturing temporal dependencies in the data.

\section{Material \& Methods}

\subsection{Data set}

We analyzed the vim-2 data set~\cite{Nishimoto2014}, which was originally published by~\cite{Nishimoto2011}. The experimental procedures are identical to those in~\cite{Nishimoto2011}. Briefly, the data set has twelve 600 s blocks of stimulus and response sequences in a training set and nine 60 s blocks of stimulus and response sequences in a test set. The stimulus sequences are videos (128 px $\times$ 128 px or 20$^\circ$ $\times$  20$^\circ$, 15 FPS) that were drawn from various sources. The response sequences are BOLD responses (voxel size = 2 $\times$ 2 $\times$ 2.5 mm$^3$, TR = 1 s) that were acquired from the occipital cortices of three subjects (S1, S2 and S3). The stimulus sequences in the test set were repeated ten times. The corresponding response sequences were averaged over the repetitions. The response sequences have already been preprocessed as described in~\cite{Nishimoto2011}. Briefly, they have been realigned to compensate for motion, detrended to compensate for drift and z-scored. Additionally, the first six seconds of the blocks were discarded. No further preprocessing was performed. Regions of interests were localized using the multifocal retinotopic mapping technique on retinotopic mapping data that were acquired in separate sessions~\cite{Hansen2004}. As a result, the voxels were grouped into 16 areas. However, not all areas were identified in all subjects. The last 45 seconds of the blocks in the training set were used as the validation set. Data were analyzed over individual subjects. Results were reported over all subjects.

\begin{figure}[h]
\begin{center}
\includegraphics[width=\textwidth]{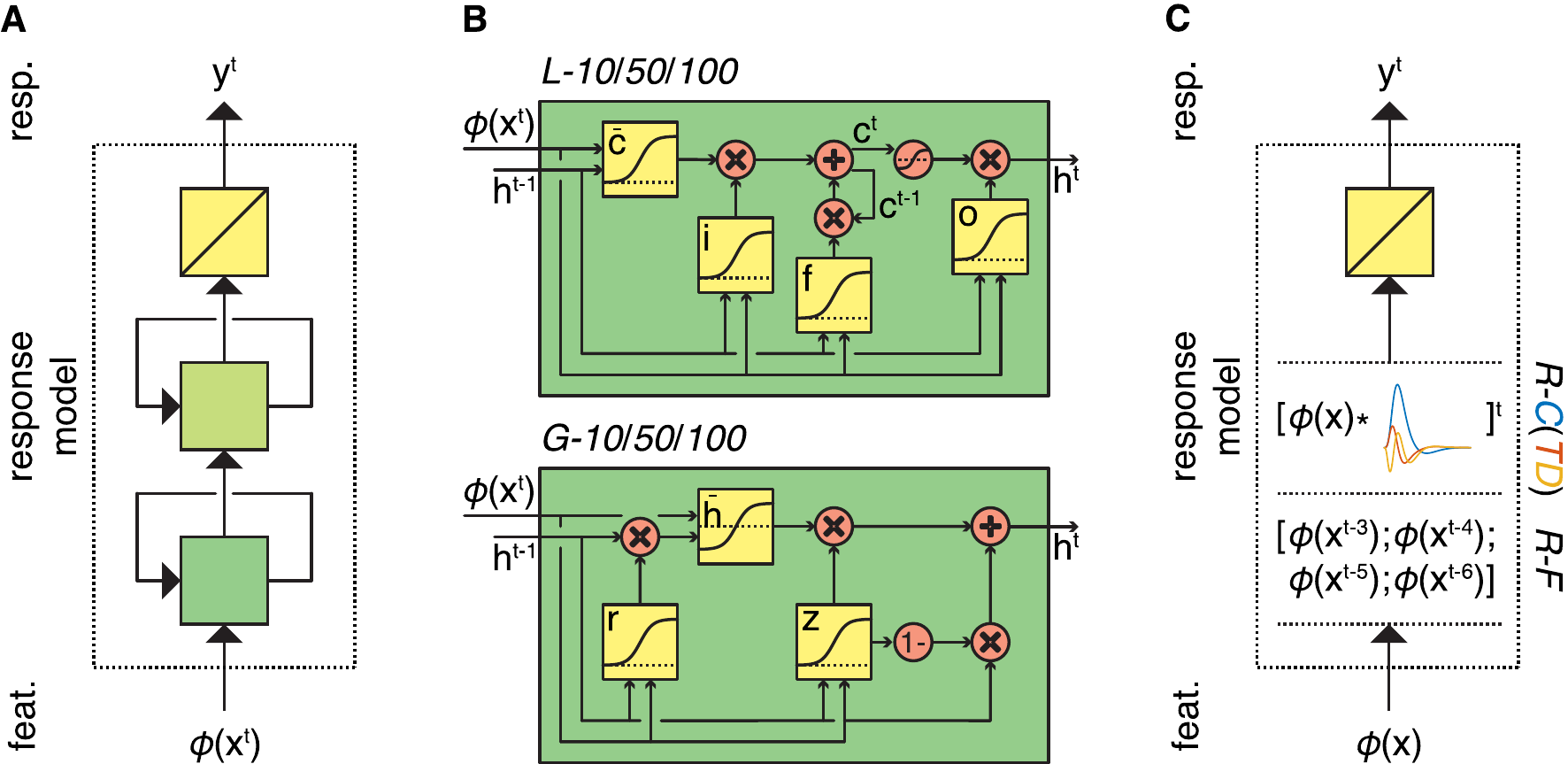}
\end{center}
 \textbf{\refstepcounter{figure}\label{fig:01} Figure \arabic{figure}.}{ \textbf{Overview of the response models.}
(\textbf{A}) Response models in the RNN family. All RNN models process feature sequences via two (recurrent) nonlinear layers and one (nonrecurrent) linear layer but differ in the type and number of artificial neurons. \textit{L-10/50/10} models have 10, 50 or 100 long short-term memory units in both of their hidden layers, respectively. Similarly, \textit{G-10/50/10} models have 10, 50 or 100 gated recurrent units in both of their hidden layers, respectively.
(\textbf{B}) First-layer long short-term memory and gated recurrent units. Squares indicate linear combination and nonlinearity. Circles indicate elementwise (non)linearity. Gates in the units control the information flow between the time points.
(\textbf{C}) Response models in the ridge regression family. All ridge regression models process feature sequences via one (nonrecurrent) linear layer but differ in how they account for the hemodynamic delay. \textit{R-C(TD)} models convolve the feature sequence with the canonical hemodynamic response function (and its time and dispersion derivatives). \textit{R-F} model lags the feature sequence for 3, 4, 5 and 6 seconds and concatenates the lagged sequences. }
\end{figure}

\subsection{Problem statement}

Let $\mathbf{x}^t\in\mathds{R}^n$ and $\mathbf{y}^t \in\mathds{R}^m$ be a stimulus and a response at temporal interval $[t,t+1]$, where $n$ is the number of stimulus dimensions and $m$ is the number of voxel responses. We are interested in predicting the most likely response $\mathbf{y}^t$ given the stimulus history $\mathbf{X}^t = (\mathbf{x}^0,\ldots,\mathbf{x}^t)$:
\begin{eqnarray}
\hat{\mathbf{y}}^t&=&\arg \max_{\mathbf{y}^t}\Pr\left(\mathbf{y}^t\middle|\mathbf{X}^t\right)\\
&=&\g\left(\f\left(\mathbf{x}^0\right),\ldots,\f\left(\mathbf{x}^t\right)\right)
\end{eqnarray}
where $\Pr$ is an encoding distribution, $\f$ is a feature model such that $\f\left(\cdot \right)\in\mathds{R}^p$, $p$ is the number of feature dimensions, and $\g$ is a response model such that $\g\left( \cdot \right)\in\mathds{R}^{m}$.

In order to solve this problem, we must define the feature model that transforms stimuli to features and the response model that transforms features to responses. We used two alternative feature models; a scene description model that codes for low-level visual features~\cite{Oliva2001} and a word embedding model that codes for high-level semantic content. We used two response model families that differ in architecture (recurrent neural network family and feedforward ridge regression family) (Figure~\ref{fig:01}).

\subsection{Feature models}

\subsubsection{High-level semantic model}

As a high-level semantic model we used the word2vec (W2V) model by~\cite{Mikolov2013a,Mikolov2013b,Mikolov2013c}. This is a one-layer feedforward neural network that is trained for predicting either target words/phrases from source-context words (continuous bag-of-words) or source context-words from target words/phrases (skip-gram). Once trained, its hidden states are used as continuous distributed representations of words/phrases. These representations capture many semantic regularities. We used the pretrained (skip-gram) W2V model to avoid training from scratch (\url{https://code.google.com/archive/p/word2vec/}). It was trained on 100 billion-word Google News dataset. It contains 300-dimensional continuous distributed representations of three million words/phrases.

We used the W2V model for transforming a stimulus sequence to a feature sequence on a second-by-second basis as follows: First, each one second of the stimulus sequence is assigned 20 categories (words/phrases). We used the \textit{Clarifai} API (\url{http://www.clarifai.com/}) to automatically assign the categories. Then, each category is transformed into continuous distributed representations of words/phrases. Next, these representations are averaged over the categories. This resulted in a 300-dimensional feature vector per second of stimulus sequence ($p=300$).

\subsubsection{Low-level visual feature model}

As a low-level visual feature model we used the GIST model~\cite{Oliva2001}. The GIST model transforms scenes into spatial envelope representations. These representations capture many perceptual dimensions that represent the dominant spatial structure of a scene and have been used to study neural representations in a number of earlier work \cite{Leeds2013,Groen2013,Cichy2016}. We used the implementation that is provided at: \url{http://people.csail.mit.edu/torralba/code/spatialenvelope/}.

We used the GIST model for transforming a stimulus sequence to a feature sequence on a second-by-second basis as follows: First, each 16 non-overlapping 8 $\times$ 8 regions of all 15 128 $\times$ 128 frames in one second of the stimulus sequence are filtered with 32 Gabor filters that have eight orientations and four scales. Then, their energies are averaged over the frames. This resulted in a 512-dimensional feature vector per second of stimulus sequence ($p=512$).

\subsection{Response models}

\subsubsection{Ridge regression family}

The response models in the ridge regression family predict feature-evoked responses as a linear combination of features. Each member of this family differs in how it accounts for the hemodynamic delay.

The \textit{R-C} model (i) convolves the features with the canonical hemodynamic response function \cite{Friston1994} and (ii) predicts the responses as a linear combination of these features:
\begin{eqnarray}
\hat{\mathbf{y}}^t&=&\left(\mathbf{H}_c\mathbf{F}_c\mathbf{B}^\top\right)^t
\end{eqnarray}
where $\mathbf{H}_c\in\mathds{R}^{t\times t}$ is the Toeplitz matrix of the canonical HRF. Furthermore, $\mathbf{F}_c=\left[\f\left(\mathbf{x}^0\right),\ldots,\f\left(\mathbf{x}^t\right)\right]^\top\in\mathds{R}^{t\times p}$ and $\mathbf{B}\in\mathds{R}^{m\times p}$ is the matrix of regression coefficients.

The \textit{R-CTD} model (i) convolves the features with the canonical hemodynamic response function, its temporal derivative and its dispersion derivative \cite{Friston1998}, (ii) concatenates these features and (iii) predicts the responses as a linear combination of these features:
\begin{eqnarray}
\hat{\mathbf{y}}^t&=&\left(\left[\mathbf{H}_c\mathbf{F}_c,\mathbf{H}_{ct}\mathbf{F}_c,\mathbf{H}_{cd}\mathbf{F}_c\right]\mathbf{B}^\top\right)^t
\end{eqnarray}
where $\mathbf{H}_{ct}\in\mathds{R}^{t\times t}$ is the Toeplitz matrix of the the temporal derivative of the canonical HRF, $\mathbf{H}_{cd}\in\mathds{R}^{t\times t}$ is the Toeplitz matrix of the the dispersion derivative of the canonical HRF and $\mathbf{B}\in\mathds{R}^{m\times 3p}$ is the matrix of regression coefficients.

The \textit{R-F} model is a finite impulse response (FIR) model that (i) lags the features for 3, 4, 5 and 6 seconds \cite{Nishimoto2011}, (ii) concatenates these features and (iii) predicts the responses as a linear combination of these features: 
\begin{eqnarray}
\hat{\mathbf{y}}^t&=&\mathbf{F}_f\mathbf{B}^\top
\end{eqnarray}
where $\mathbf{F}_f=\left[\f\left(\mathbf{x}^{t-3}\right),\f\left(\mathbf{x}^{t-4}\right),\f\left(\mathbf{x}^{t-5}\right),\f\left(\mathbf{x}^{t-6}\right)\right]^\top\in\mathds{R}^{t\times 4p}$ and $\mathbf{B}\in\mathds{R}^{m\times 4p}$ is the matrix of regression coefficients.

We used the validation set for model selection (a regularization parameter per voxel) and the training set for model estimation (a row of $\mathbf{B}$ per voxel). Regularization parameters were selected as explained in \cite{Guclu2014}. The rows of $\mathbf{B}$ were estimated by analytically minimizing the $L^2$-penalized least squares loss function.

\subsubsection{Recurrent neural network family}

The response models in the RNN family are two-layer recurrent neural network models. They use their internal memories for nonlinearly processing arbitrary feature sequences and predicting feature-evoked responses as a linear combination of their second-layer hidden states:
\begin{eqnarray}
\hat{\mathbf{y}}^t=\mathbf{h}_2^t\mathbf{W}^\top
\end{eqnarray}
where $\mathbf{h}_2^t$ represents the hidden states in the second layer, and $\mathbf{W}$ are the weights. The RNN models differ in the type and number of artificial neurons.

The \textit{L-10}, \textit{L-50} and \textit{L-100} models are two-layer recurrent neural networks that have 10, 50 and 100 long short-term memory (LSTM) units \cite{Hochreiter1997} in the their hidden layers, respectively. The first-layer hidden states of an LSTM unit are defined as follows:
\begin{eqnarray}
\mathbf{h}^t&=&\mathbf{o}^t\odot\tanh\left(\mathbf{c}^t\right)\\
\mathbf{o}^t&=&\sigma\left(\mathbf{U}_o\mathbf{h}^{t-1}+\mathbf{W}_o\f\left(\mathbf{x}^t\right)+\mathbf{b}_o\right)
\end{eqnarray}
where $\odot$ denotes elementwise multiplication, $\mathbf{c}^t$ is the cell state, and $\mathbf{o}^t$ are the output gate activities. The cell state maintains information about the previous time points. The output gate controls what information will be retrieved from the cell state. The cell state of an LSTM unit is defined as:
\begin{eqnarray}
\mathbf{c}^t&=&\mathbf{f}^t\odot\mathbf{c}^{t-1}+\mathbf{i}^t\odot\bar{\mathbf{c}}^t\\
\mathbf{f}^t&=&\sigma\left(\mathbf{U}_f\mathbf{h}^{t-1}+\mathbf{W}_f\f\left(\mathbf{x}^t\right)+\mathbf{b}_f\right)\\
\mathbf{i}^t&=&\sigma\left(\mathbf{U}_i\mathbf{h}^{t-1}+\mathbf{W}_i\f\left(\mathbf{x}^t\right)+\mathbf{b}_i\right)\\
\bar{\mathbf{c}}^t&=&\sigma\left(\mathbf{U}_c\mathbf{h}^{t-1}+\mathbf{W}_c\f\left(\mathbf{x}^t\right)+\mathbf{b}_c\right)
\end{eqnarray}
where $\mathbf{f}^t$ are the forget gate activities, $\mathbf{i}^t$ are the input gate activities, and $\bar{\mathbf{c}}^t$ is an auxiliary variable. Forget gates control what old information will be discarded from the cell states. Input gates control what new information will be stored in the cell states. The remaining variables that are not explicitly described are the model parameters.

The \textit{G-10}, \textit{G-50} and \textit{G-100} models are two-layer recurrent neural networks that have 10, 50 and 100 gated recurrent units (GRU) \cite{Cho2014} in the their hidden layers, respectively. The GRU units are simpler alternatives to the LSTM units. They combine hidden states with cell states and input gates with forget gates. The first-layer hidden states of a GRU unit is defined as follows:
\begin{eqnarray}
\mathbf{h}^t&=&\left(1- \mathbf{z}^t\right)\odot\mathbf{h}^{t-1}+\mathbf{z}^t\odot\bar{\mathbf{h}}^t\\
\mathbf{z}^t&=&\sigma\left(\mathbf{U}_z\mathbf{h}^{t-1}+\mathbf{W}_z\f\left(\mathbf{x}^t\right)+\mathbf{b}_z\right)\\
\mathbf{r}^t&=&\sigma\left(\mathbf{U}_r\mathbf{h}^{t-1}+\mathbf{W}_r\f\left(\mathbf{x}^t\right)+\mathbf{b}_r\right)\\
\bar{\mathbf{h}}^t&=&\tanh\left(\mathbf{U}_h\left(\mathbf{r}_t\odot\mathbf{h}^{t-1}\right)+\mathbf{W}_h\f\left(\mathbf{x}^t\right)+\mathbf{b}_h\right)
\end{eqnarray}
where $\mathbf{z}^t$ are update gate activities, $\mathbf{r}^t$ are reset gate activities and $\bar{\mathbf{h}}^t$ is an auxiliary variable. Like the gates in LSTM units, those in GRU units control the information flow between the time points. The remaining variables that are not explicitly described are the parameters of the model.

The second-layer hidden states are defined similarly to the first-layer hidden states except for replacing the input features with the first-layer hidden states). We used truncated backpropagation through time in conjunction with Adam \cite{Kingma2014} to train the models on the training set. Dropout \cite{Hinton2012} was used to regularize the hidden layers. The epoch in which the validation performance was the highest was taken as the final model. The \textit{Chainer} framework (\url{http://chainer.org/}) was used to implement the models.

\subsection{HRF estimation}

Voxel-specific HRFs were estimated by stimulating the RNN model with an impulse. Let $\mathbf{x}^{-t}, \ldots, \mathbf{x}^0, \ldots, \mathbf{x}^t$ be an impulse such that $\mathbf{x}$ is a vector of zeros at times other than time 0 and a vector of ones at time 0. The period of the impulse before time 0 is used to stabilize the baseline of the impulse response. First, the response of the model to the impulse is simulated:
\begin{eqnarray}
\left[\mathbf{H}_r^*\right]_{-t}^t&=&\g_r\left(\mathbf{x}^{-t},\ldots,\mathbf{x}^0,\ldots,\mathbf{x}^t\right)
\end{eqnarray}
where $\left[\mathbf{H}_r^*\right]_{-t}^t=\left(\mathbf{H}_r^{*-t},\ldots,\mathbf{H}_r^{*0},\ldots,\mathbf{H}_r^{*t}\right)$. Then, the baseline of the impulse response before time 0 is subtracted from itself:
\begin{eqnarray}
\left[\mathbf{H}_r^*\right]_{-t}^t&=&\left[\mathbf{H}_r^*\right]_{-t}^t-\mathbf{H}_r^{*-1} \,.
\end{eqnarray}
Next, the impulse response is divided by its maximum:
\begin{eqnarray}
\left[\mathbf{H}_r^*\right]_{-t}^t&=&\left[\mathbf{H}_r^*\right]_{-t}^t / \max\left[\mathbf{H}_r^*\right]_{-t}^t \,.
\end{eqnarray}
Finally, the period of the impulse response before time 0 is discarded, and the remaining period of the impulse response is taken as the HRF of the voxels:
\begin{eqnarray}
\left[\mathbf{H}_r\right]_0^t&=&\left[\mathbf{H}_r^*\right]_{0}^t \,.
\end{eqnarray}

The time when the HRF is at its maximum was taken as the delay of the response, and the time after the delay of the response when the HRF was at its minimum was taken as the delay of undershoot.

\section{Results}

The performance of a model in a voxel was defined as the Pearson's product-moment correlation coefficient between the observed and predicted responses of the voxel ($r$). Its performance in a group of voxels was defined as the median of its performance over the voxels in the group ($\tilde{r}$).

In order to make sure that the differences in the performance of a model in different areas are not caused by the differences in the signal-to-noise ratios of the areas, the performance of the model in an area was corrected for the median of the noise ceilings of the voxels in the area ($\tilde{r}^*$) \cite{Kay2013}.

\subsection{Comparison of the response models}

We evaluated the response models by comparing the performance of the response models in the (recurrent) RNN family and (feed-forward) ridge regression family in combination with the (high-level) W2V model and the (low-level) GIST model. Using two feature models of different levels ruled out any potential biases in the performance difference of the response models that can be caused by the feature models. Recall that the models in the RNN family (\textit{G/L-10}/\textit{50}/\textit{100} models) differed in the type and number of artificial neurons, whereas the models in the ridge regression family (\textit{R-C}/\textit{R-CTD}/\textit{R-F} models) differed in how they account for the hemodynamic delay.

Once the best response models among the RNN family and the ridge regression family were identified, we first compared their performance in detail. Particular attention was paid to the voxels where the performance of the models differed by more than an arbitrary threshold of $r=0.1$. We then compared the performance of the best response model among the RNN family over the areas along the visual pathway.

\subsubsection{Comparison of the response models in combination with the semantic model}

\begin{figure}
\begin{center}
\includegraphics[width=\textwidth]{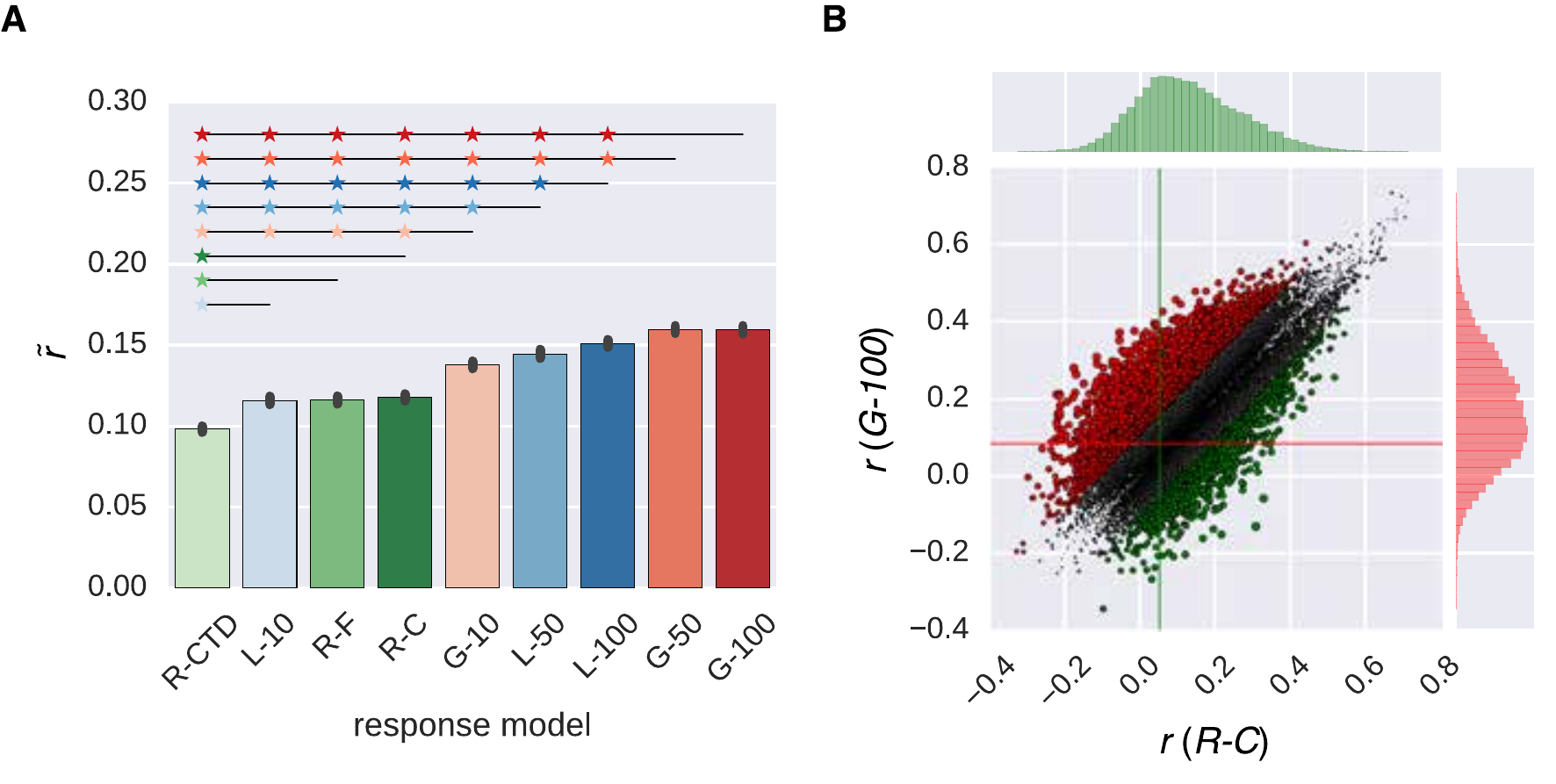}
\end{center}
 \textbf{\refstepcounter{figure}\label{fig:02} Figure \arabic{figure}.}{ \textbf{Comparison of the response models in combination with the W2V model.} (\textbf{A}) Median performance of response models in RNN (\textit{G-X} and \textit{L-X}) and ridge regression (\textit{R-X}) families over all voxels. Error bars indicate 95\% confidence intervals (bootstrapping). Asterisks indicate significant performance difference. (\textbf{B}) Performance of best response models in RNN (\textit{G-100} model) and ridge regression (\textit{R-C} model) families over individual voxels. Points indicate voxels. Gray points indicate voxels where the performance difference is more than $r=0.1$. Lines indicate (median) performance over all voxels. }
\end{figure}

Figure~\ref{fig:02} compares the performance of all response models in combination with the W2V model. The performance of the models in the RNN family that had 50 or 100 artificial neurons were always significantly higher than that of all models in the ridge regression family (p $\leq$ 0.05, bootstrapping method). However, the performance of the models in the same family was not always significantly different from each other. The performance of the \textit{G-100} model was the highest among the RNN family ($\tilde{r}=0.16$), and that of the \textit{R-C} model was the highest among the ridge regression family ($\tilde{r}=0.12$).

The performance of the \textit{G-100} model and the \textit{R-C} model differed from each other by more than the chosen threshold of $r=0.1$ in 30\% of the voxels. The performance of the \textit{G-100} model was higher in 78\% of these voxels ($\Delta\tilde{r}=0.17$), and that of the \textit{R-C} model was higher in 22\% of these voxels ($\Delta\tilde{r}=0.14$).

\begin{figure}
\begin{center}
\includegraphics[width=\textwidth]{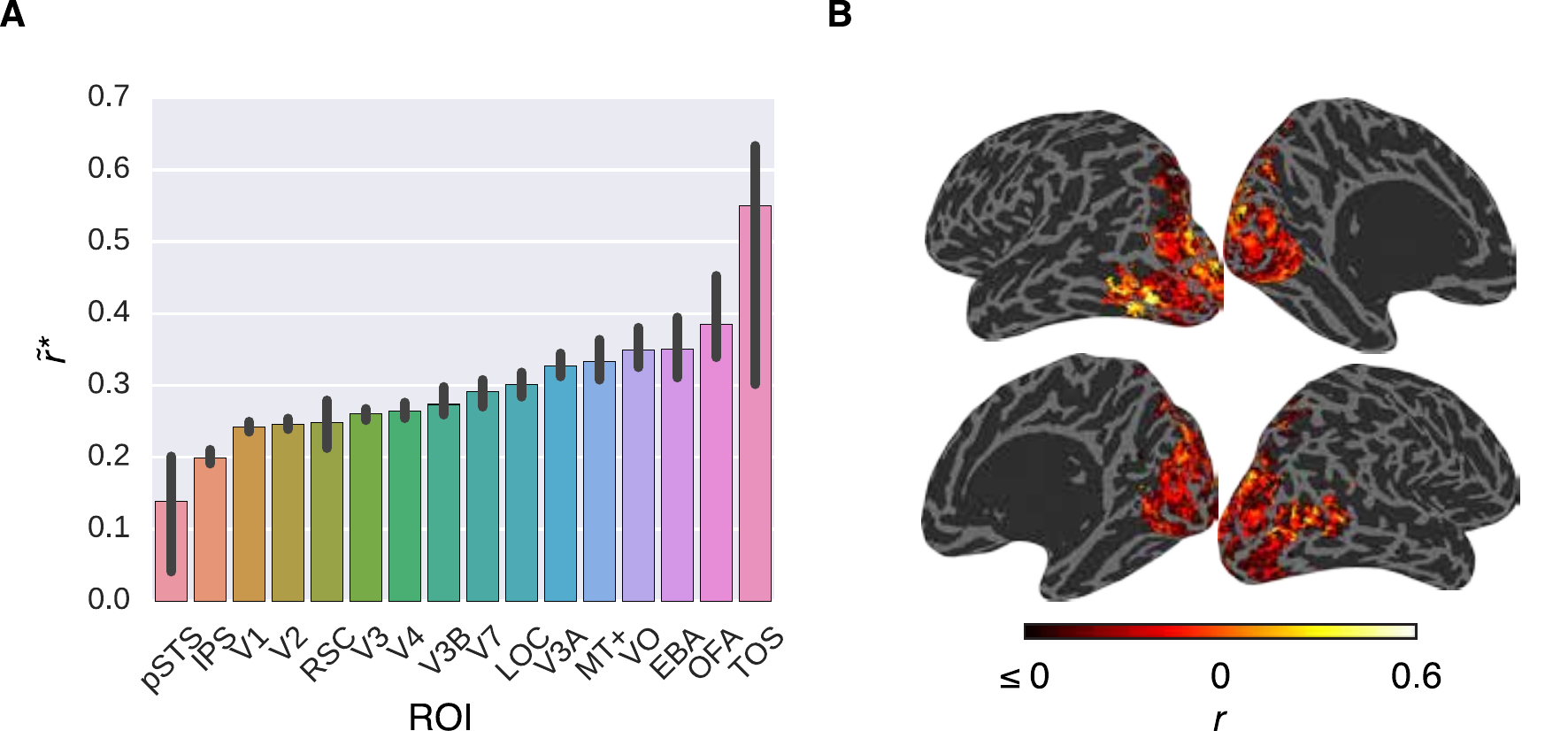}
\end{center}
 \textbf{\refstepcounter{figure}\label{fig:03} Figure \arabic{figure}.}{ \textbf{Comparison of the \textit{G-100} model in combination with the W2V model in different areas.} (\textbf{A}) Median noise ceiling controlled performance over all voxels in different areas. Error bars indicate 95\% confidence intervals (bootstrapping). (\textbf{B}) Projection of performance to cortical surfaces of S3. }
\end{figure}

Figure~\ref{fig:03} compares the performance of the \textit{G-100} model in combination with the W2V model over the areas along the visual stream. While the performance of the model was significantly higher than chance throughout the areas (p $\leq$ 0.05, bootstrapping method), it was particularly high in downstream areas. For example, it was the highest in TOS ($\tilde{r}^*=0.55$), OFA ($\tilde{r}^*=0.38$) and EBA ($\tilde{r}^*=0.35$), and the lowest in pSTS ($\tilde{r}^*=0.14$), IPS ($\tilde{r}^*=0.20$) and V1 ($\tilde{r}^*=0.24$).

\subsubsection{Comparison of the response models in combination with the low-level feature model}

\begin{figure}
\begin{center}
\includegraphics[width=\textwidth]{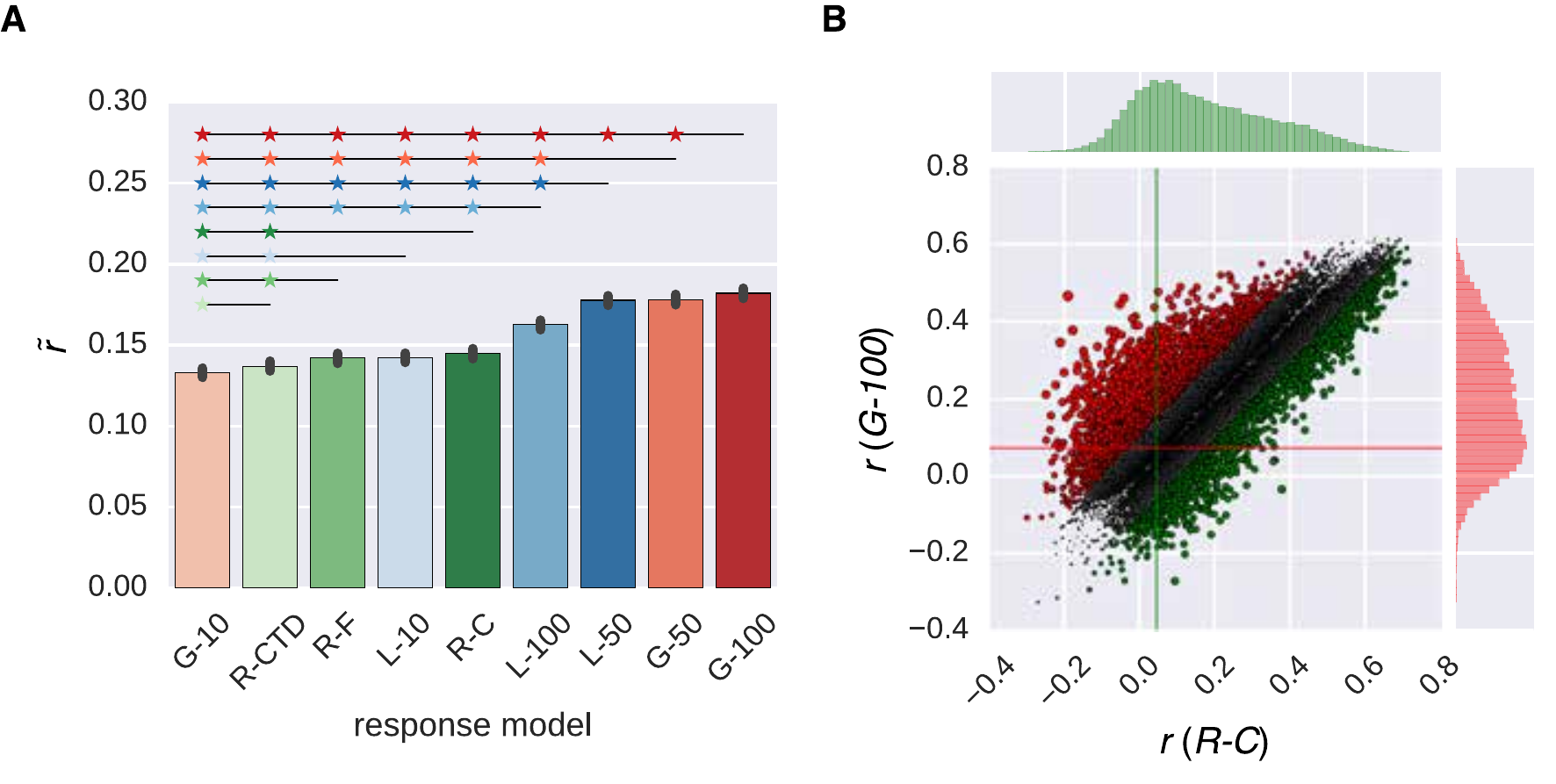}
\end{center}
 \textbf{\refstepcounter{figure}\label{fig:04} Figure \arabic{figure}.}{ \textbf{Comparison of the response models in combination with the GIST model.} (\textbf{A}) Median performance of response models in RNN (\textit{G-X} and \textit{L-X}) and ridge regression (\textit{R-X}) families over all voxels. Error bars indicate 95\% confidence intervals (bootstrapping). Asterisks indicate significant performance difference. (\textbf{B}) Performance of best response models in RNN (\textit{G-100} model) and ridge regression (\textit{R-C model}) families over individual voxels. Points indicate voxels. Gray points indicate voxels where the performance difference is more than $r=0.1$. Lines indicate median performance over all voxels. }
\end{figure}

Figure~\ref{fig:04} compares the performance of the all response models in combination with the GIST model. The trends that were observed in this figure were similar to those that were observed in Figure~\ref{fig:02}. The \textit{G-100} model was the best among the RNN family ($\tilde{r}=0.18$), and the \textit{R-C} model was the best among the ridge regression family ($\tilde{r}=0.14$).

The \textit{G-100} model and the \textit{R-C} differed from each other by more than the threshold of $r=0.1$ in 27\% of the voxels. The \textit{G-100} model was better in 66\% of these voxels ($\Delta\tilde{r}=0.17$). The \textit{R-C} model was better in 34\% of these voxels ($\Delta\tilde{r}=0.14$).

\begin{figure}
\begin{center}
\includegraphics[width=\textwidth]{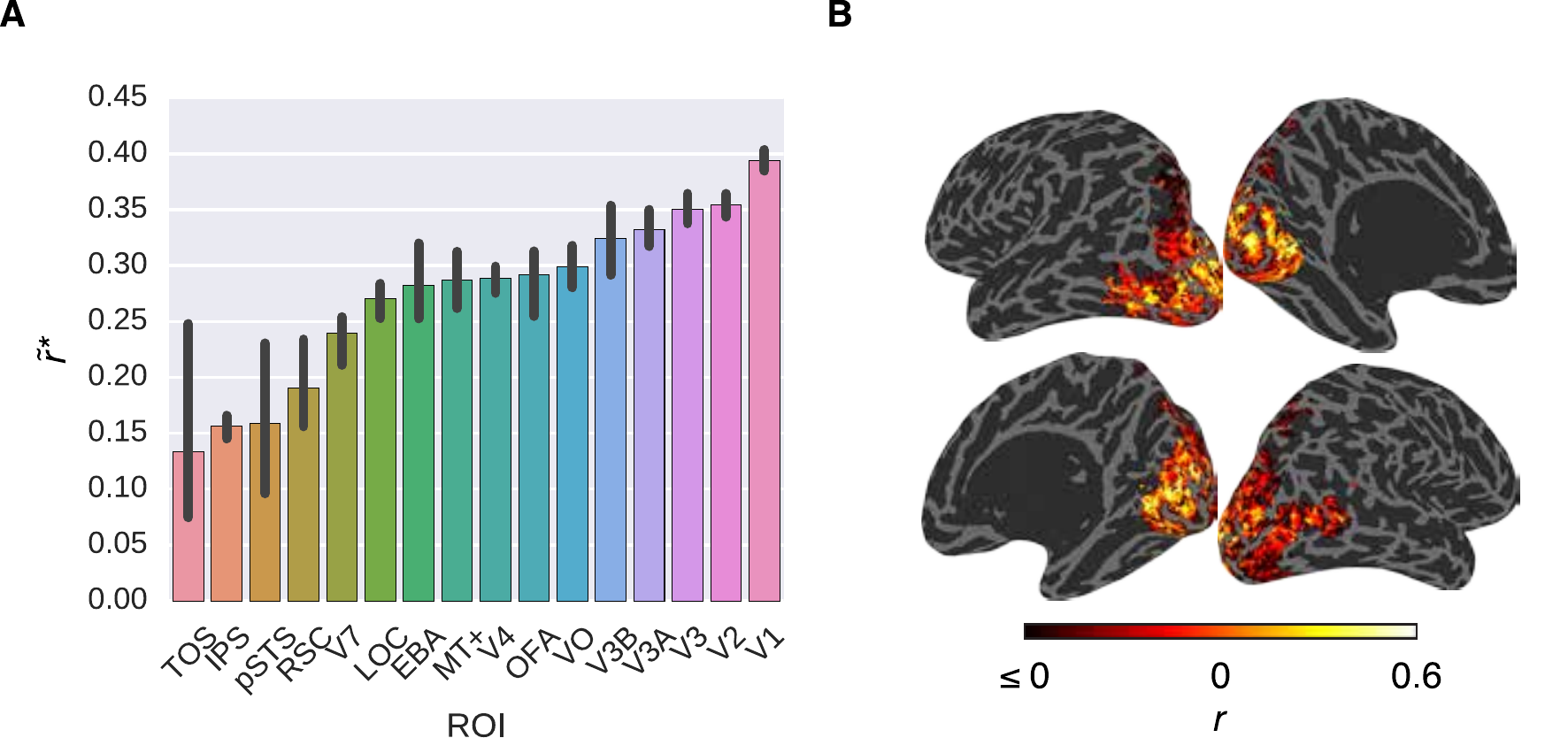}
\end{center}
 \textbf{\refstepcounter{figure}\label{fig:05} Figure \arabic{figure}.}{ \textbf{Comparison of the \textit{G-100} model in combination with the GIST model in different areas.} (\textbf{A}) Median noise ceiling controlled performance over all voxels in different areas. Error bars indicate 95\% confidence intervals (bootstrapping). (\textbf{B}) Projection of performance to cortical surfaces of S3. }
\end{figure}

Figure~\ref{fig:05} compares the performance of the \textit{G-100} model in combination with the GIST model over the areas along the visual pathway. While the \textit{G-100} model performed significantly better than chance throughout the areas ($p\leq0.05$, bootstrapping method), it performed particularly well in upstream visual areas. For example, it performed the best in V1 ($\tilde{r}^*=0.39$), V2 ($\tilde{r}^*=0.35$) and V3 ($\tilde{r}^*=0.35$), and the worst in TOS ($\tilde{r}^*=0.13$), IPS ($\tilde{r}^*=0.16$) and pSTS ($\tilde{r}^*=0.16$).

\subsection{Comparison of the feature models}

\begin{figure}
\begin{center}
\includegraphics[width=\textwidth]{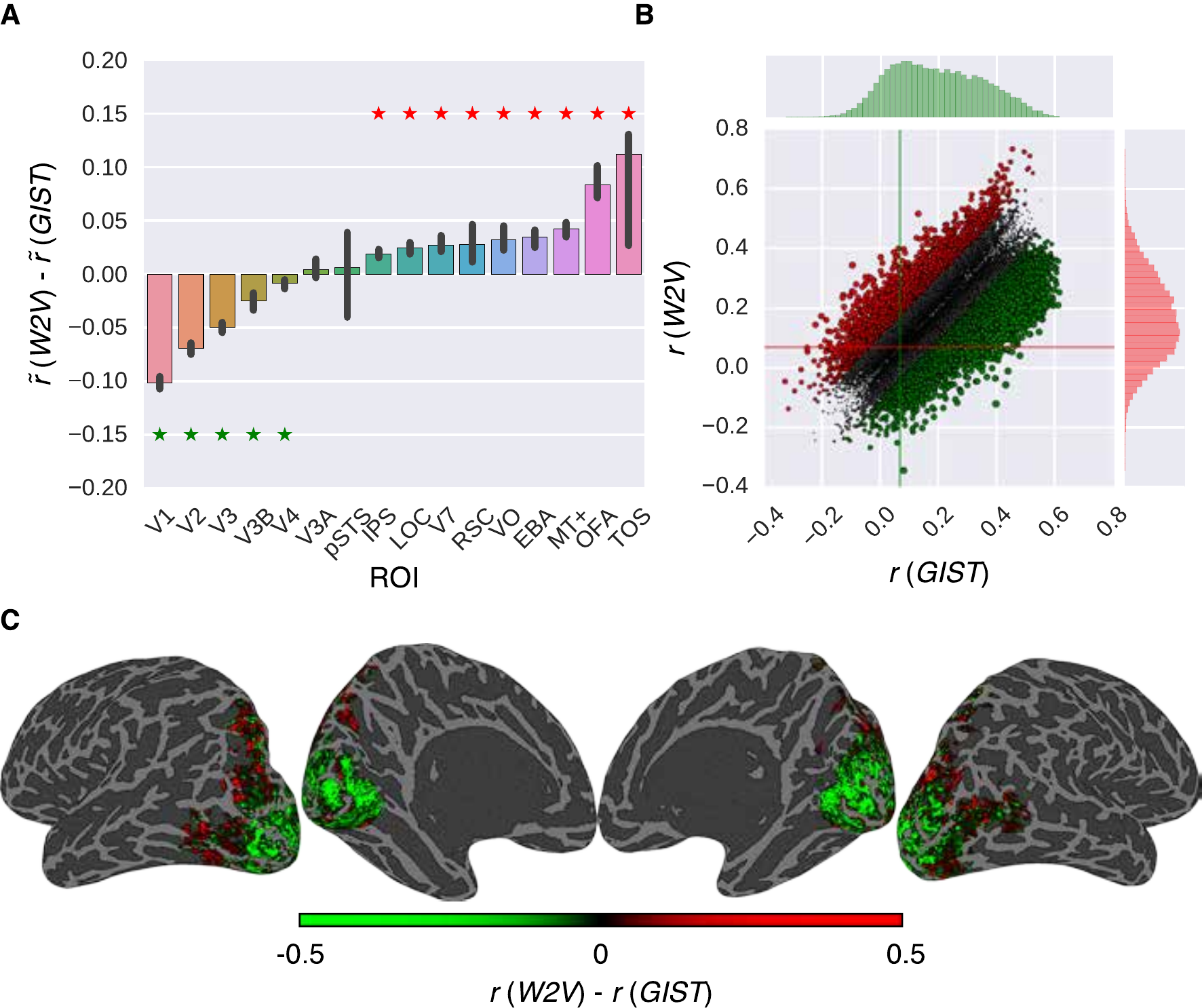}
\end{center}
 \textbf{\refstepcounter{figure}\label{fig:06} Figure \arabic{figure}.}{ \textbf{ Comparison of the feature models in combination with the \textit{G-100} model.} (\textbf{A}) Median performance difference over all voxels in different areas. Asterisks indicate significant performance difference. Error bars indicate 95\% confidence intervals (bootstrapping). (\textbf{B}) Performance over individual voxels. Points indicate voxels. Gray points indicate voxels where performance difference is more than $r=0.1$. Lines indicate median performance over all voxels. (\textbf{C}) Projection of performance difference to cortical surfaces of S3. }
\end{figure}

Once the efficacy of the proposed RNN models was positively assessed, we assessed the extent to which the voxels prefer semantic representations versus perceptual representations by comparing the performance of the W2V model and the GIST model in combination with the \textit{G-100} model (Figure~\ref{fig:06}).

The performance of the models was significantly different in all areas along the visual stream except for pSTS and V3A ($p \leq 0.05$, bootstrapping method). This difference was in favor of semantic representations in downstream areas and perceptual representations in upstream areas. The largest difference in favor of semantic representations was in TOS ($\Delta\tilde{r}=0.11$), OFA ($\Delta\tilde{r}=0.08$) and MT+ ($\Delta\tilde{r}=0.04$), and perceptual representations was in V1 ($\Delta\tilde{r}=0.10$), V2 ($\Delta\tilde{r}=0.07$) and V3 ($\Delta\tilde{r}=0.05$).

Thirty-nine percent of the voxels preferred either representation by more than the arbitrary threshold of $r=0.1$. Thirty-four percent of these voxels preferred semantic representations ($\Delta\tilde{r}=0.16$), and 66\% percent of these voxels preferred perceptual representations ($\Delta\tilde{r}=0.18$).

\subsection{Estimation of voxel-specific HRFs}

\begin{figure}
\begin{center}
\includegraphics[width=\textwidth]{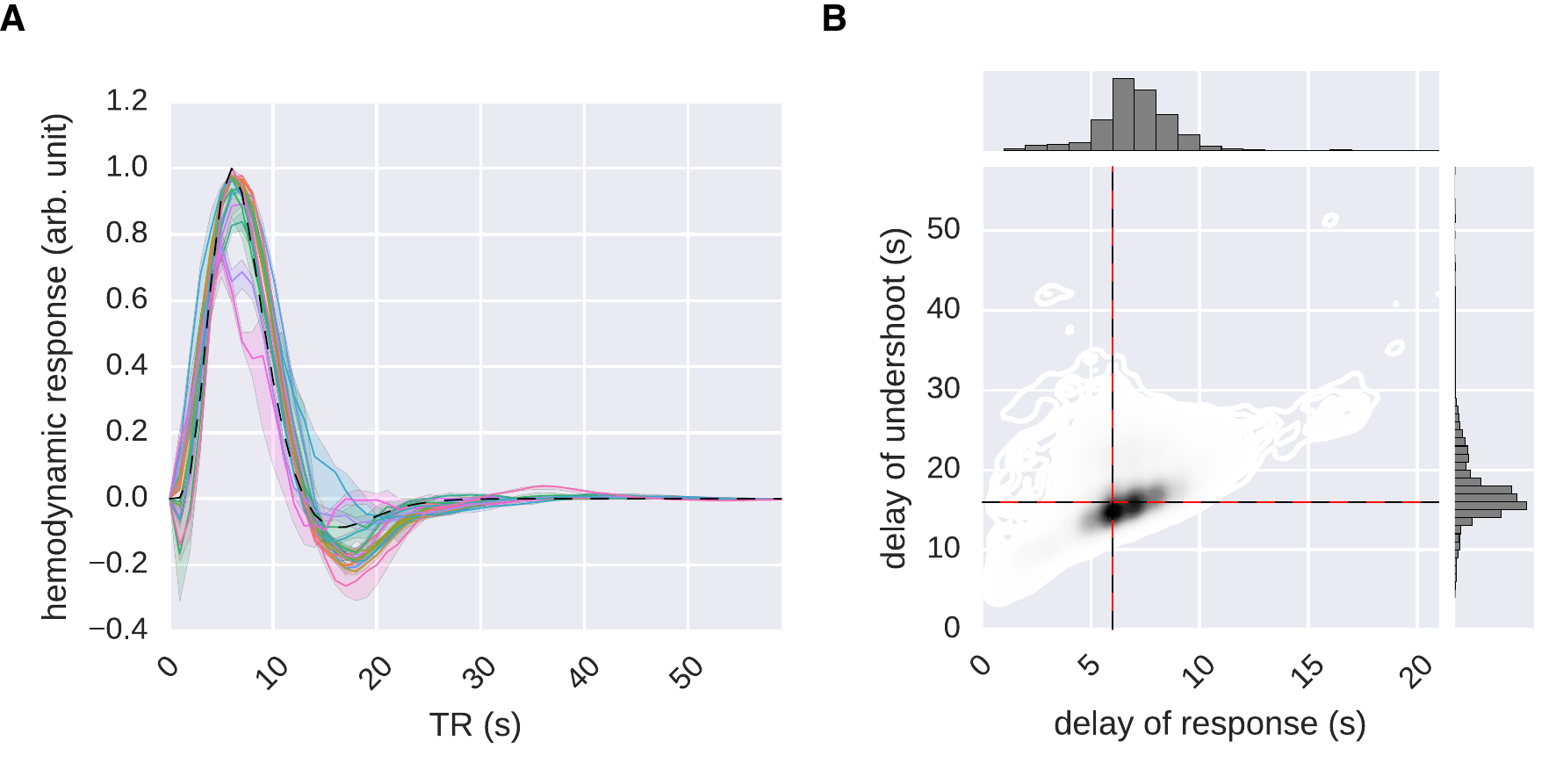}
\end{center}
 \textbf{\refstepcounter{figure}\label{fig:07} Figure \arabic{figure}.}{ \textbf{Estimation of the hemodynamic response functions.} The \textit{G-100} model was stimulated with an impulse. The impulse response was processed by normalizing its baseline and scale. The result was taken as the HRF. (\textbf{A}) Median hemodynamic response functions of all voxels in different areas. Error bands indicate 68\% confidence intervals (bootstrapping). Different colors indicate different areas. Dashed line indicate canonical hemodynamic response function. (\textbf{B}) Delays of responses and undershoots of all voxels. Red lines indicate median estimated delays over all voxels. Black line indicates canonical delays. }
\end{figure}

One important advantage of the response models in the RNN family is that they can not only learn (stimulus) feature-response relationships but also estimate HRFs of voxels. We used both feature models in combination with the \textit{G-100} model to estimate the HRFs of the voxels where the performance of any model combination was significantly higher than chance (51\% of the voxels, $p\leq0.05$, Student's t-test, Bonferroni correction) (Figure~\ref{fig:07}). The W2V and \textit{G-100} models were used to estimate the HRFs of the voxels where their performance was higher than that of the GIST and \textit{G-100} models, and vice versa. 

While the global shape of the estimated HRFs was similar to that of the canonical HRF, there was a considerable spread in the estimated delays of responses and the delays of undershoots (median delay of response = 6.57 $\pm$ 0.02 s, median delay of undershoot = 16.95 $\pm$ 0.04 s). Furthermore, the delays of responses were significantly correlated with the delays of undershoots (Pearson's $r=0.45$, $p\leq0.05$, Student's $t$-test).

\subsection{Analysis of the internal representations of the \textit{G-100} model}

\begin{figure}
\begin{center}
\includegraphics[width=\textwidth]{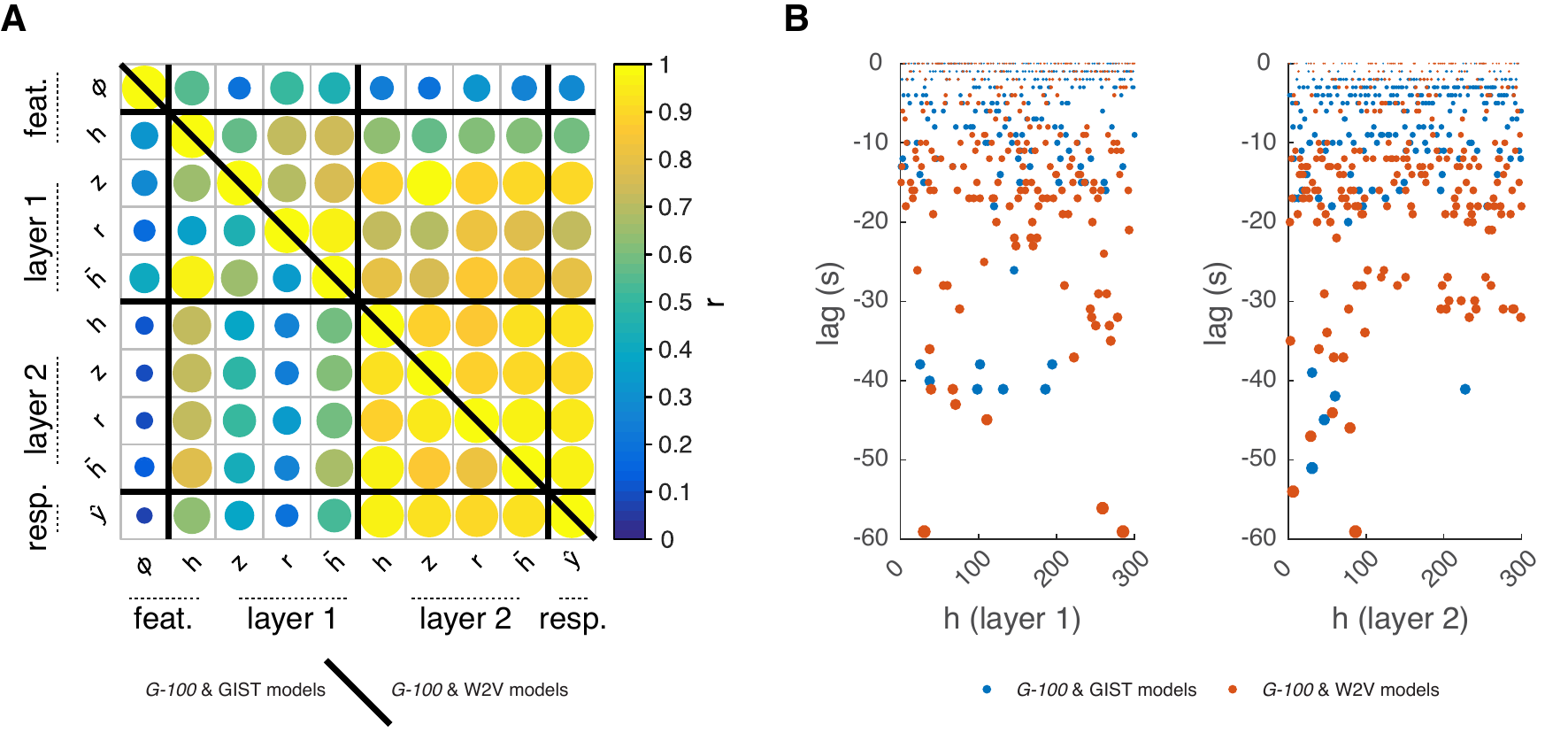}
\end{center}
 \textbf{\refstepcounter{figure}\label{fig:08} Figure \arabic{figure}.}{ \textbf{Internal representations of the \textit{G-100} model.} (\textbf{A}) Overlap between representational dissimilarity matrices of features, layer 1 internal states, layer 2 internal states and predicted responses. Upper and lower triangular parts correspond to W2V and GIST models, respectively.
(\text{B}) Temporal selectivity of layer 1 and layer 2 hidden units. Points indicate lags at which cross-correlations between hidden states and features are highest.}
\end{figure}

Finally, to gain insight into other temporal dependencies captured by the \textit{G-100} model, we analyzed its internal representations (Figure~\ref{fig:08}).

We first constructed representational dissimilarity matrices (RDMs) of the stimulus sequence in the test set at different stages of the processing pipeline and averaged them over subjects \cite{Kriegeskorte2008}. Per feature model, this resulted in one RDM for the features, four RDMs for the internal states in a layer and one RDM for the predicted responses. We then correlated the upper triangular parts of the RDMs with one another. We found that the RDMs of the internal states (in particular those in layer 2) were highly correlated with the predicted response RDM but less so with the feature RDMs. The mean correlations between the feature and internal state RDMs were 0.39 for layer 1 and 0.21 for layer 2. Those between the internal state and predicted response RDMs were 0.61 for layer 1 and 0.93 for layer 2.

Next, each hidden state was cross-correlated with each stimulus feature, and the cross-correlations were averaged over the features. This resulted in a value that shows how much a hidden state is selective to the features at different time points. The time point at which this value was at its maximum was taken as the optimal lag of that hidden unit. W

\section{Discussion}

Understanding how the human brain responds to its environment is a key objective in neuroscience. This study has shown that recurrent neural networks are exquisitely capable of capturing how brain responses are induced by sensory stimulation, outperforming established approaches based on ridge regression. This increased sensitivity has important consequences for future studies in this area. One might argue that the non-linearities introduced by the use of RNNs make them harder to interpret compared to the use of a linear mapping. However, when the goal is to compare between different computational models, such as the GIST and W2V models used here, maximizing explained variance becomes the main criterion of interest.

Furthermore, RNNs that make use of gated recurrent units were shown to outperform long short-term memory units. However, this is likely to be caused by difficulties in model estimation in the current data regime rather than one RNN architecture being better suited to the problem at hand than the other.

\subsection{Testing hypotheses about brain function}

Like any other encoding model, RNN based encoding models can be used to test hypotheses about neural representations \cite{Naselaris2011}. That is, they can be used to test whether a particular feature model outperforms alternative feature models when it comes to explaining observed data. As such, we have shown that a low-level visual feature model explains responses in upstream visual areas well, whereas a high-level semantic model explains responses in downstream visual areas well. This complements existing work on deep neural networks (DNNs), where a similar dissociation between upstream and downstream visual areas can be observed by mapping different DNN layers to different brain regions~\cite{Guclu2015a}. 

Furthermore, RNN-based encoding models can also be used to test hypotheses about the temporal dependencies between features and responses. For example, by constraining the temporal memory capacities of the RNN units, one can identify the optimal scale of the temporal dependencies that different brain regions are selective to.

\subsection{Capturing temporal dependencies}

RNNs can use their internal memories to capture the temporal dependencies in data. In the context of modeling the dynamics of brain activity in response to naturalistic stimuli, these dependencies can be caused by either neurovascular coupling or by stimulus-induced cognitive processes.

By providing an RNN with an impulse on the input side, it was shown that, effectively, the RNN learns to represent voxel-specific hemodynamic responses. Importantly, the RNNs allowed us to estimate these HRFs from data collected under a continuous design. To the best of our knowledge this is the first time it has been shown that this is possible in practice. 

Results also show that RNN's internal representations are strongly correlated with predicted responses but less so with stimulus features. This suggests that the RNN carries information from features at the time points in the past to make predictions on the responses at the time points in the present. As such, we have shown that the internal representations of RNNs are selective for different periods in the feature timecourses that span a large temporal range. This is likely induced by stimulus-related and cognitive factors on top of the hemodynamic response. Yet, it  remains a challenge to interpret the remaining temporal dependencies that are learned by RNNs. Combining the present work with recent developments on understanding RNN representations \cite{Karpathy2015} might shed further light on these dependencies.

\subsection{Other uses of RNNs}

Here, we used RNNs as response models in an encoding framework. That is, they were used to predict responses to features that were extracted from stimuli with separate feature models. However, use cases of RNNs are not limited to this setting. There are a number of other ways in which RNNs can be used to understand how sensory information is represented in the brain.

First, RNN models can be used as feature models instead of response models in the encoding framework. Like CNNs, RNNs are being used to solve various problems in fields ranging from computer vision~\cite{Gregor2015} to computational linguistics~\cite{Zaremba2014}. Internal representations of task-optimized CNNs were shown to correspond to neural representations in different brain regions~\cite{Kriegeskorte2015,Yamins2016a}. It would be interesting to see if the internal representations of task-based RNNs have similar correlates in the brain.

Second, RNN models can be used as an alternative to the standard multi-voxel pattern analysis (MVPA) models~\cite{Haynes2015} in the decoding framework. In this setting, MVPA models are used for classifying or regressing stimuli from stimulus-evoked responses. Instead of raw responses, response amplitudes that are estimated for each trial and voxel combination by deconvolving voxel-specific response time courses from the time series data with a general linear model are typically fed into the MVPA models. This means that the MVPA models might also suffer from the problems that are caused by temporal dependencies. Therefore, we expect RNNs to outperform also the standard MVPA models in the decoding framework.

\subsection{Limitations of RNNs}

RNNs can process arbitrary input sequences in theory. However, they have an important limitation in practice. Like any other contemporary neural network architecture, typical RNN architectures have a very large number of free parameters. Therefore, a very large amount of training data is required for accurately estimating RNN models without overfitting. While there are several methods to combat overfitting in RNNs like different variants of dropout \cite{Hinton2012,Zaremba2014,Semeniuta2016}, it is still an important issue to which particular attention needs to be paid. This also means that RNN models will face difficulties when trying to predict responses to very high-dimensional stimulus features such as the internal representations of convolutional neural networks which range from thousands to hundreds of thousands dimensions. Therefore, for such features, dimensionality reduction techniques need to be utilized for reducing the feature dimensionality to a range that can be handled with RNNs in scenarios with either insufficient computational resources or training data.

\subsection{Conclusions}

We have shown for the first time that RNNs can be used to predict how the human brain processes sensory information. Whereas classical connectionist research has focused on the use of RNNs as models of cognitive processing~\cite{Elman1993}, the present work has shown that RNNs can also be used to probe the neural correlates of ongoing cognitive processes induced by dynamically changing naturalistic sensory stimuli. The ability of RNNs to learn about long-range temporal dependencies provides the flexibility to couple ongoing sensory stimuli that induce various cognitive processes with delayed measurements of brain activity that depend on such processes. This end-to-end training approach can be applied to any neuroscientific experiment in which sensory inputs are coupled to observed neural responses.

\clearpage

\bibliographystyle{ieeetr}
\bibliography{main}

\end{document}